\documentclass[11pt]{article}
\usepackage{txfonts}
\usepackage{cospar}
\usepackage[sectionbib]{natbib}
\renewcommand{\bibsection}{\section*{REFERENCES}}
\renewcommand{\bibhang}{\setlength{8mm}}
\renewcommand{\bibsep}{\setlength{0mm}}
\usepackage{graphicx}

\newcommand\src{G21.5--0.9}

\newcommand\E[1]{\times10^{#1}}
\newcommand\U[1]{\,{\mathrm{#1}}}

\newcommand\rs[1]{_{\rm #1}}

\newcommand\tauscaone{\tau\rs{sca,1}}
\newcommand\amax{a\rs{max}}
\newcommand\NH{N\rs{H}}

\newcommand\sg{\sigma}

\title{THE X-RAY HALO OF G21.5--0.9}

\author{R. Bandiera\address{Osservatorio Astrofisico di Arcetri,
        Largo E. Fermi 5, I-50125, Firenze, Italy},
        and
        F. Bocchino\address{Osservatorio Astronomico ``G.S. Vaiana'',
        Piazza del Parlamento 1, I-90134, Palermo, Italy}}

\begin{document}

\maketitle

\begin{abstract}
The emission of the plerion \src\ appears more extended in X rays than in
radio. This is an unexpected result because it would imply that short-lived
X-ray electrons may reach distances even larger than radio electrons. Applying
an empirical relationship between dust scattering optical depth and
photoelectric column density, the measured column density leads to a large
optical depth at 1~keV, of about 1. Therefore we investigate the hypothesis
that the detected halo be an effect of dust scattering, re-analyzing an Cal/PV
XMM-Newton observation of \src\ and critically examining it in terms of a dust
scattering model. We also present a spectral analysis of a prominent extended
feature in the northern sector of the halo.
\end{abstract}

\section*{INTRODUCTION}
\vspace{3pt}

In various respects supernova remnant \src\ looks like a standard plerion. In
fact it shows: i.\ a filled-center morphology, both in radio and in X rays;
ii.\ a flat radio synchrotron spectrum; iii.\ a non-thermal X-ray spectrum,
with photon index $\sim2$. There is no direct evidence of the pulsar powering
the nebula (no pulsed emission has been detected as yet), but \src\ harbours an
almost point-like central X-ray source, that has been interpreted as the
termination shock of the pulsar wind (Slane et al., 2000).

However a non-thermal X-ray halo has been recently detected, extending well
beyond the boundaries of the radio emitting plerion (Warwick et al., 2001).
This is the opposite of what observed in other plerions, as well as of what
standard models of plerions would predict. Our main goal is to investigate the
origin of this halo.

Moreover, a result of past X-ray spectral analyses of the halo region (Warwick
et al., 2001; Safi-Harb et al., 2001) has been that the overall spectrum of the
halo does not show any thermal signature. No thermal emission has been even
detected in ``the North Spur'', a feature in the northern part of the halo,
that morphologically resembles a partial shell. In this paper we shall also
discuss this point.

\vspace{6pt}
\section*{THE NATURE OF THE X-RAY HALO}
\vspace{3pt}
\subsection*{Difficulties with the standard interpretation}

An X-ray synchrotron nebula with a size larger than the radio emitting nebula
represents a challenge to the existing models of plerion. In fact the
synchrotron emitting relativistic electrons are expected to be injected in the
immediate surroundings of the associated pulsar, and then to be advected and/or
diffused to fill the whole nebula; in the meanwhile synchrotron losses are
selectively quenching higher energy particles. Therefore X-ray emitting
electrons should typically disappear before reaching the boundary of the
magnetic bubble, which instead should be rather well outlined at radio
wavelengths. The above scenario nicely explains (at least at a qualitative
level) the behavior of most plerions, in which the apparent size of the nebula
is shrinking with increasing photon energy.

In order to account for the opposite behavior, as observed in \src,  one may
for instance invoke higher diffusion for higher energy electrons. But then, in
order to effectively get synchrotron emission, strong magnetic fields are
required: therefore the magnetic bubble should extend to the whole X-ray halo;
but in this case it is hard to explain why the radio emitting electrons do not
fill the whole magnetic bubble. More detailed modelling may be devised, but the
above reasoning should have made clear the theoretical embarrassment rising
after the \src\ case.

\subsection*{An alternative explanation: dust scattering}

An alternative interpretation is that the plerionic nebula is not intrinsically
more extended in X rays than in radio; but that the observed X-ray halo is just
an effect of dust scattering in the foreground medium.

Following the empirical relation derived by Predehl and Schmitt (1995), the
measured column density $\NH=2.3\E{22}\U{cm^{-2}}$ (as from Warwick et al.,
2001) corresponds to $\tauscaone=1.07$ (where $\tauscaone$ indicates the dust
scattering optical depth at 1~keV). For such a high value the expected
fractional halo intensity is compatible with what measured (38\%) in the
0.1--2.4~keV band. However the Predehl and Schmitt statistical relation allows
a considerable dispersion, and therefore it is not sufficient by itself to
prove that dust scattering is the main cause of the halo observed in \src.

We should add that fits to the halo spectrum imply $\NH$ values significantly
lower than in the inner part of the plerion (Warwick et al., 2001; Safi-Harb et
al., 2001). This decrease of $\NH$ with increasing radius cannot be intrinsic;
it may be regarded, instead, as a sign that dust scattering distorts to some
extent the local spectra, biasing the spectral fits. Scattering may partly
affect even the observed radial softening in the X-ray halo. The analysis
described below is intended to explore in depth the dust scattering hypothesis,
by investigating the halo properties in various X-ray energy bands, taking
advantage of energy scaling relations intrinsic of the scattering physics.

\vspace{6pt}
\section*{MODELLING A DUST SCATTERING HALO}
\vspace{3pt}

\subsection*{The data}

We have reanalyzed EPIC/PN data obtained in a XMM-Newton Cal/PV observation of
\src. In Figure 1 the source is imaged in three different bands, 0.5--2, 2--5
and 5--8~keV. Overlaid to the soft image are the selection areas for the
spectral analysis described in the last section: an annulus with inner and
outer radius of respectively $70''$ and $110''$ has been subdivided into 4
sectors, labelled as
{\it halo1} (P.A. from $-30^\circ$ to  $60^\circ$), 
{\it halo2} (P.A. from  $60^\circ$ to $150^\circ$), 
{\it halo3} (P.A. from $150^\circ$ to $240^\circ$), 
{\it halo4} (P.A. from $240^\circ$ to $330^\circ$);
furthermore, a $45''$ diameter circle centered on the North Spur has been
labelled {\it Nspur}.

\begin{figure}
\includegraphics[width=0.3648\textwidth]{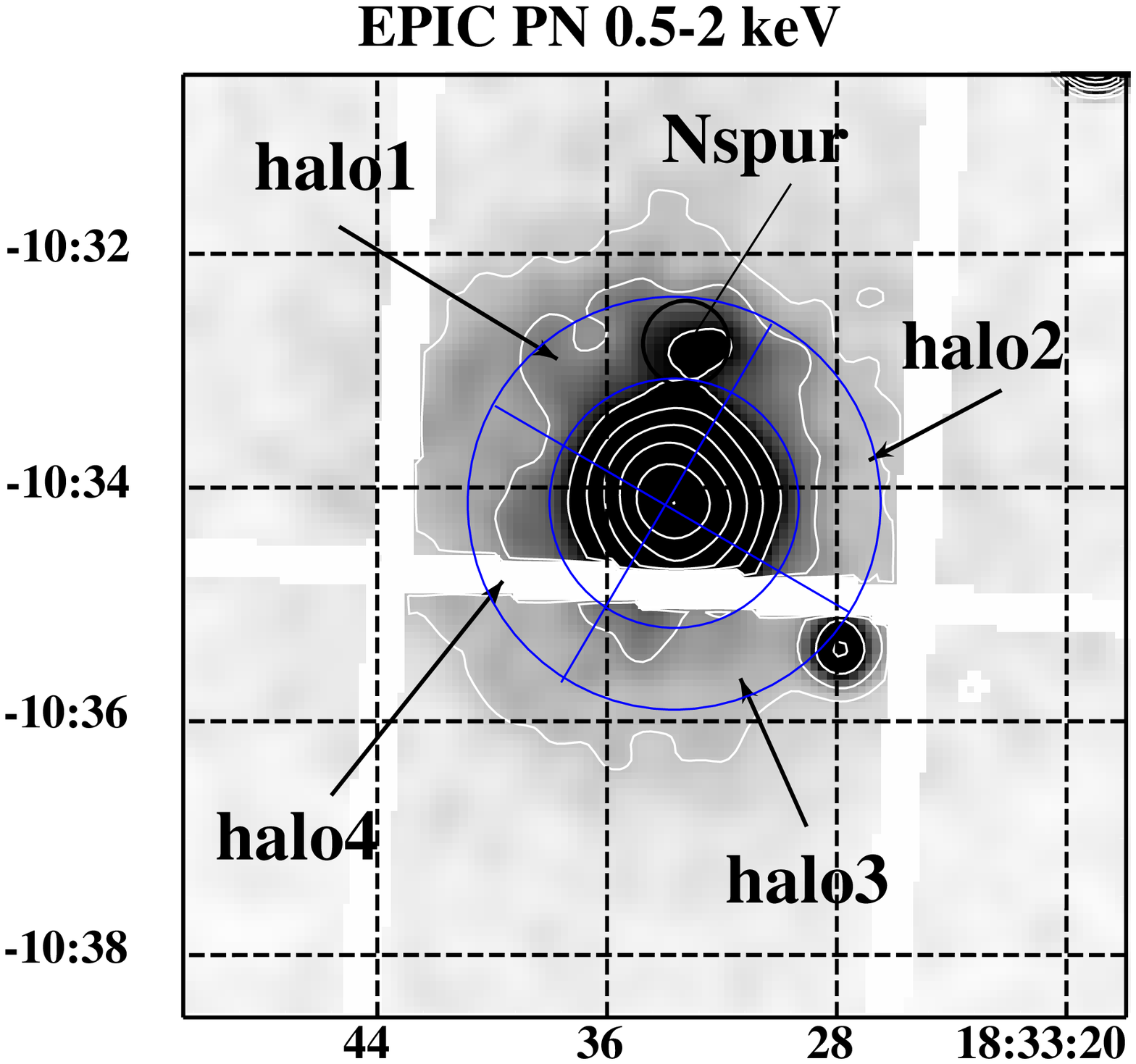}
\includegraphics[width=0.3176\textwidth]{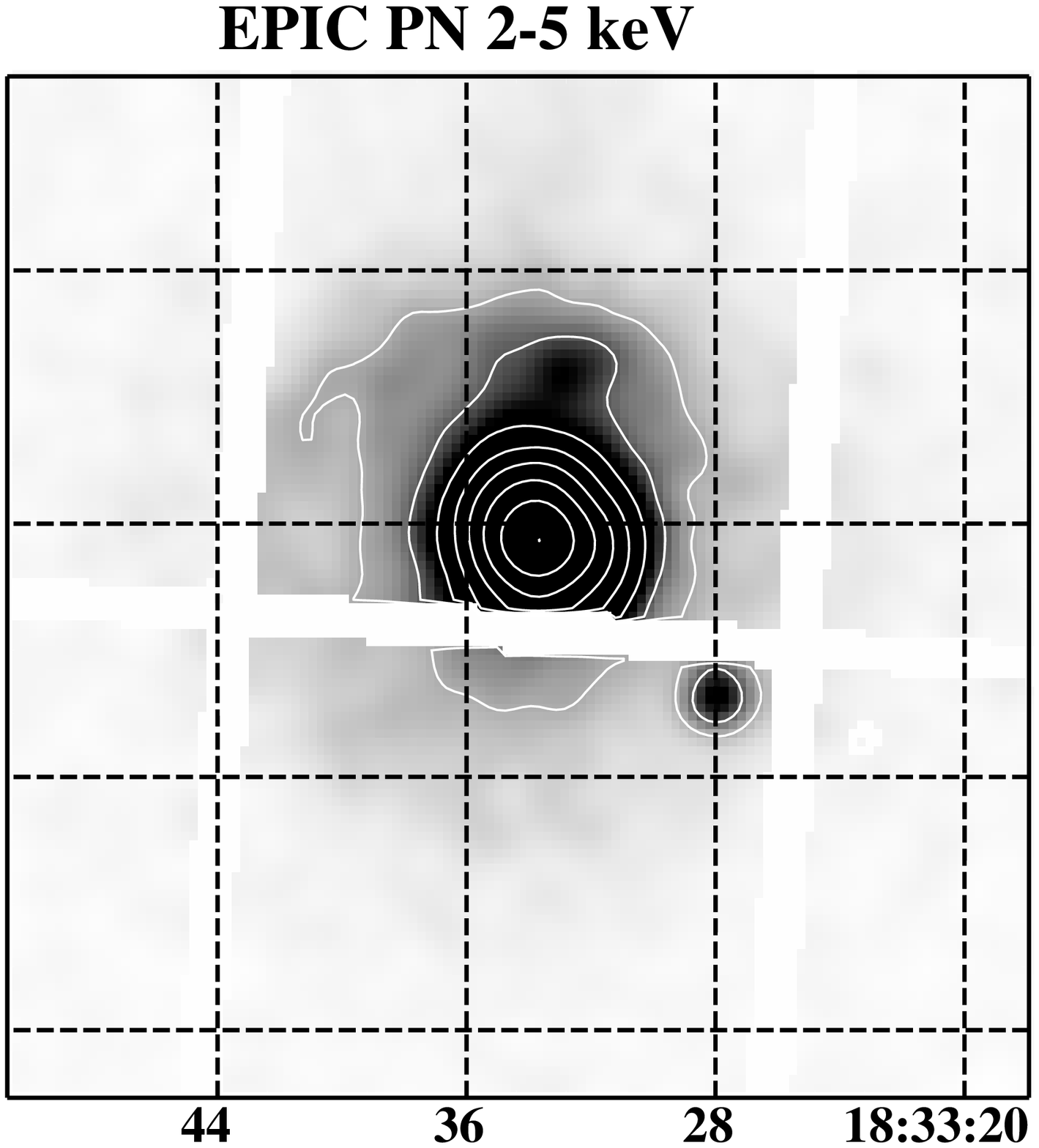}
\includegraphics[width=0.3176\textwidth]{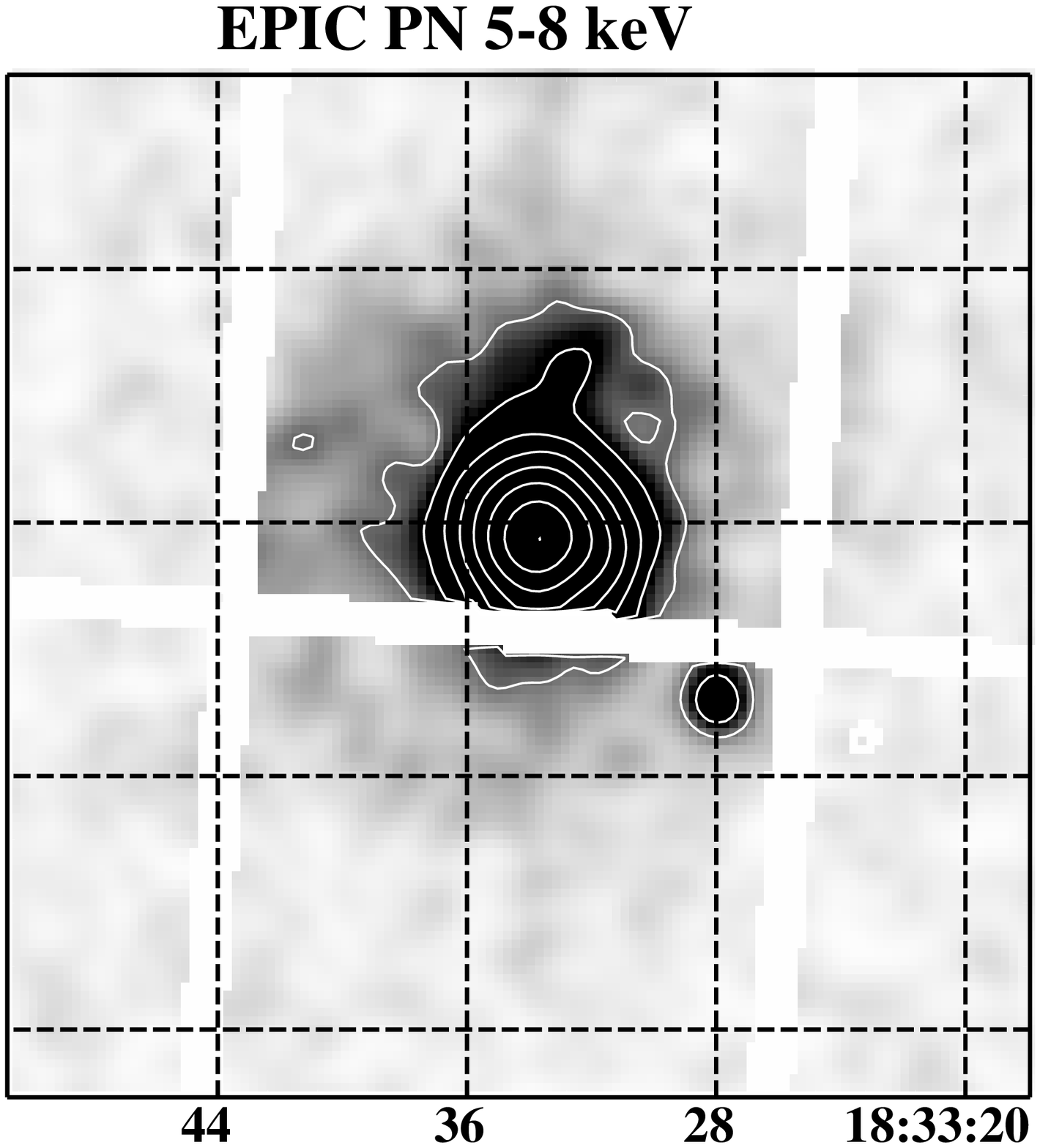}
\vspace{-24pt}
\caption{\ \
EPIC/PN images of \src\ with pixel size of $4''$ and smoothed using a 2 pixel
sigma gaussian. In the leftmost image the spectral extraction regions are
overlaid. X-ray contours are logarithmic between 1/128 and the peak value, in
steps of factor 2.
}
\end{figure}

\subsection*{The model assumptions}

Our analysis is based on some assumptions, which are listed below.

\begin{itemize}

\item We assume that the scattering is adequately described by the
Rayleigh-Gans Theory. This implies well defined energy dependences, both for
the halo scale length ($\propto E^{-1}$) and for the scattering optical depth
($\propto E^{-2}$; see e.g.\ Mauche and Gorenstein, 1986). A deviation from
these power laws is expected in soft X rays, and is actually observed in our
data (see below).

\item The dust properties (namely dust density and grain size distribution) are
taken to be homogeneous along the line of sight. In the case of variations in
the distribution, a sort of average is obtained. An uneven density distribution
along the line of sight, instead, does not affect the profile of the halo tail
(on which we base our analysis).

\item We model the intrinsic X-ray size of the source as only weakly dependent
on energy. In fact in other plerions the X-ray size may slightly shrink with
increasing energy. Anyway this would not affect the dust halo tail, where the
original source can be hardly distiguished from a point-like source.

\item We assume EPIC/PN Point Spread Function (PSF) to be energy independent.
Looking at the radial profiles of a point-like source (LMC X--3; Figure 2) it
can be easily verified to be a valid approximation.

\end{itemize}

\begin{figure}
\begin{center}
\includegraphics[width=0.90\textwidth]{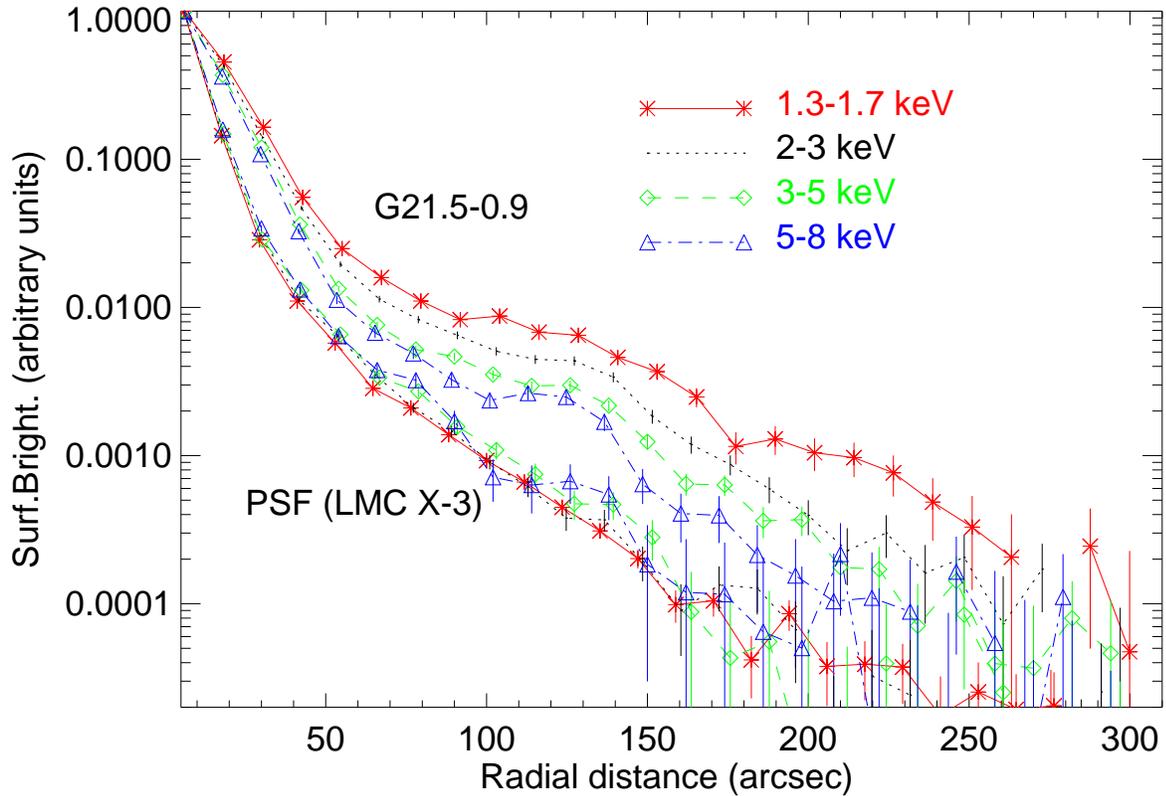}
\end{center}
\vspace{-36pt}
\caption{\ \
EPIC/PN X-ray radial profiles of \src\ and LMC X--3, in the 4 energy bands used
in this work. In \src\ radial profiles are broader at lower energies; while the
PSF is almost energy independent.
}
\end{figure}

To summarize, all above assumptions look reasonable. Moreover the final
conclusions do not strictly require them to be valid. Therefore we believe our
results are rather robust.

\subsection*{Analysis procedure}

This procedure is based on the comparison of radial profiles in various X-ray
bands. In order to obtain profiles of sufficient quality to detect even the
outermost tail, we have limited our analysis to 4 spectral bands, namely
1.3--1.7~keV, 2--3~keV, 3--5~keV, and 5--8~keV.

Figure 2 shows the 4 profiles, for \src\ as well as for LMC X--3 (a bright
point-like source with low $\NH$, used to determine the instrumental PSF). The
profiles are background subtracted and vignetting corrected, and out-of-time
events have been taken away. In the case of \src\ we have also excluded regions
of the image containing the North Spur and the star SS~397. All profiles are
shown normalized to their peak value. In the case of \src\ the radial profile
is getting broader in softer X rays: this is not an instrumental effect, since
from Figure 2 it is apparent that the PSF is energy independent till the
faintest parts of its wings.

In order to fit the dust scattering halo of \src\ we adopt a model with 3
parameters: $\tauscaone$ (optical depth at 1~keV), $\amax$ (maximum grain size)
and $q$ (power-law index of the grain distribution; Predehl and Schmitt,
1995). Taking the 5--8~keV profile \src\ as a good guess of the halo-free
source profile, a fit to the 2--3~keV source profile (see the relative curve in
Figure 3) gives $\tauscaone=1.50$, $\amax=0.90\U{\mu m}$, $q=3.3$. Then,
keeping all the best-fit parameters fixed, we have used the energy scaling laws
for dust scattering in Rayleigh-Gans approximation (size~$\propto E^{-1}$,
$\amax\propto E^{-2}$) to successfully reproduce also the profile in the
3--5~keV X-ray band, while the data at 1.3--1.7~keV look slightly overestimated
(Figure 3).

\begin{figure}
\includegraphics[width=0.5\textwidth]{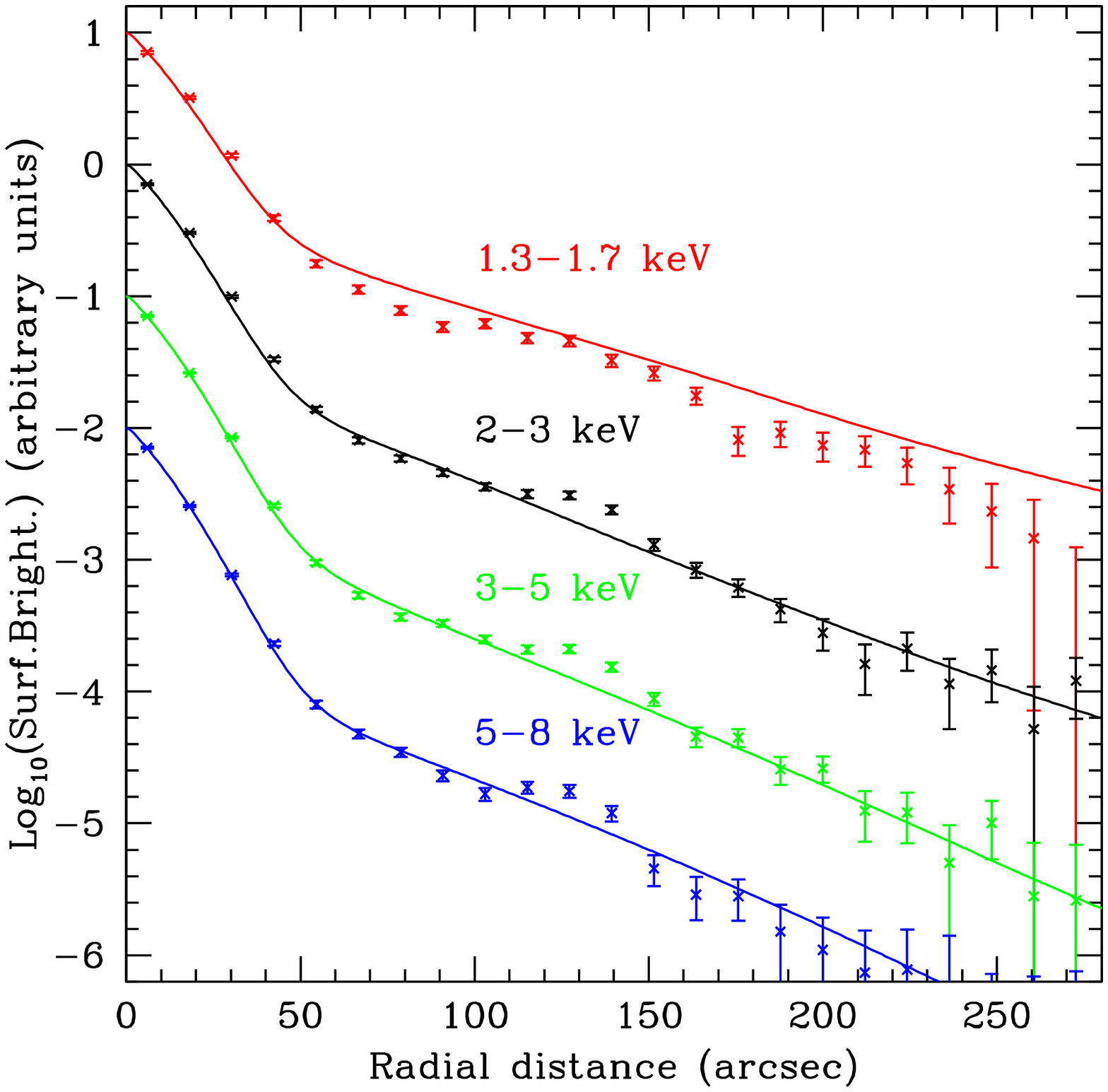}
\includegraphics[width=0.5\textwidth]{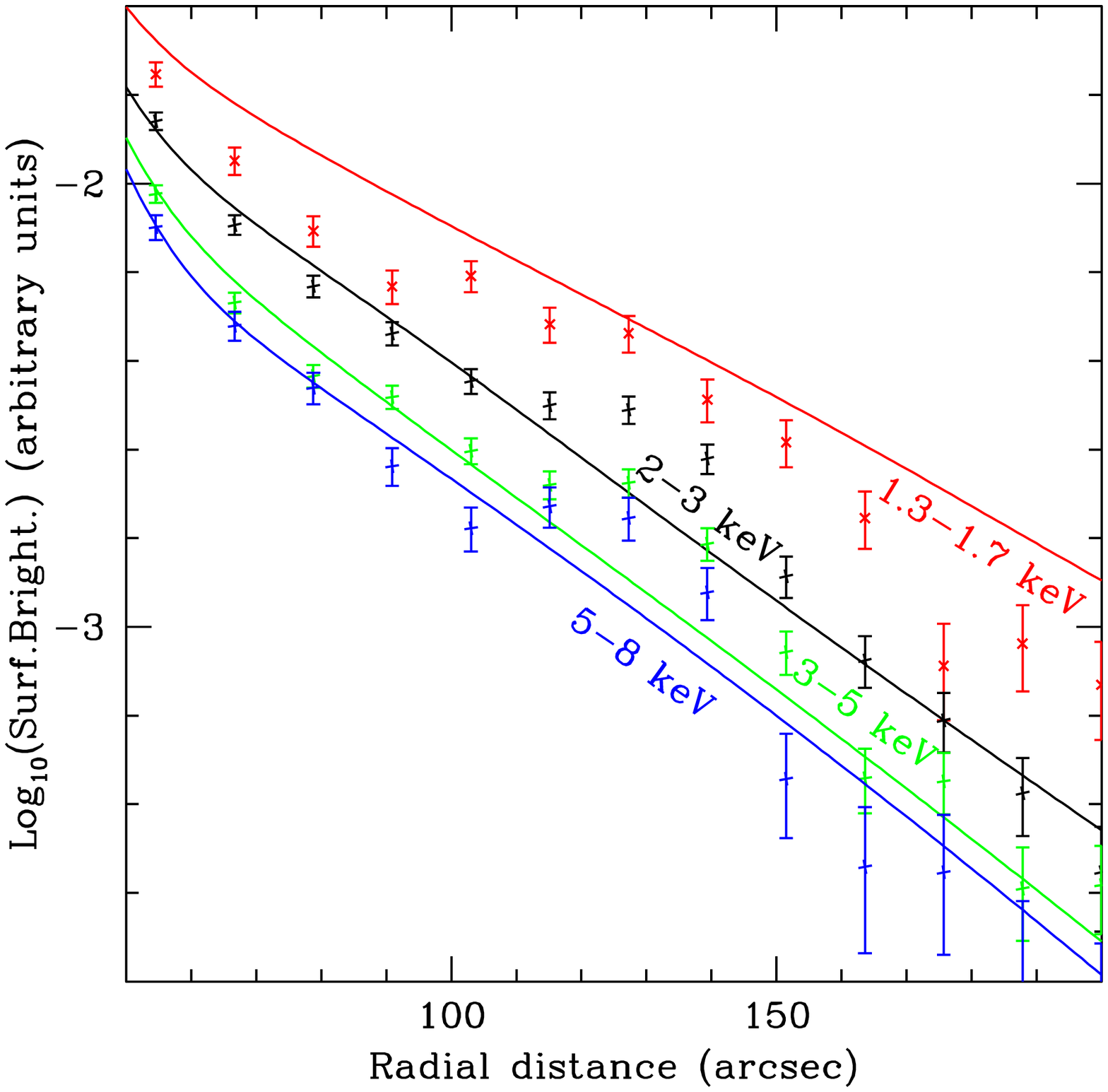}
\vspace{-24pt}
\caption{\ \
{\bf (Left)}
Dust-halo fit to the data. The model profiles are overplotted with the data.
Vertical offsets between different bands have been added for a clearer display.
{\bf (Right)}
Same as before with a different choice of the plot boundaries, in order to
better examine the behavior in the tail. The only mismatch between data and
model is in the 1.3--1.7~keV band.
}
\end{figure}

\vspace{6pt}
\section*{DISCUSSION}
\vspace{3pt}

The main results of the above analysis are the following:

\begin{itemize}

\item By comparing our best-fit $\amax$ ($0.90\U{\mu m}$) and $q$ (3.3) with
the values averaged over a large sample of sources ($\amax=0.17\U{\mu m}$ and
$q=3.9$; Predehl and Schmitt, 1995), the distribution of grain sizes turns out
to be flatter and more extended to larger sizes than the average; while, owing
to the uncertainties involved in the Predehl and Schmitt empirical relation,
the best-fit $\tauscaone$ (1.50) is in reasonable agreement with the value
(1.07) derived from $\NH$,

\item The detected slight discrepancy from the 1.3--1.7~keV data (about 30\%,
with the model overestimating the data) is naturally explained in terms of the
invalidity of the Rayleigh-Gans theory in soft X rays. Simulations with a
similar $\amax/E$ ratio (Smith and Dwek, 1998) give an offset in the same
direction as observed, ranging from 25\% (for graphite grains) to 50\% (for
silicate grains).

\item The radial profile in the 5--8 keV band, that on the basis of the above
analysis should be only slightly affected by dust scattering, shows instead a
tail broader than the PSF (Figure 4). This effect, although involving very weak
surface brightnesses ($10^{-4}$ times the peak), is significant. It may either
be an instrumental bias that we have not identified yet (and that is not
documented), or the evidence of an intrinsic extended component, even though
much less prominent than previously thought.

\item
We have tried to reproduce the data in the 5--8 keV band convolving various
simple source models with the PSF. No source smaller in size than the radio
plerion has been found to reproduce correctly the X-ray wings. A good fit is
obtained modelling the intrinsic halo as a uniform sphere of radius $200''$ and
surface brightness $10^{-4}$ times the inner component (for a total flux which
is only 2\% of that of the inner source, Figure 4), although this solution is
not unique.

\end{itemize}

\begin{figure}
\begin{center}
\includegraphics[width=0.60\textwidth]{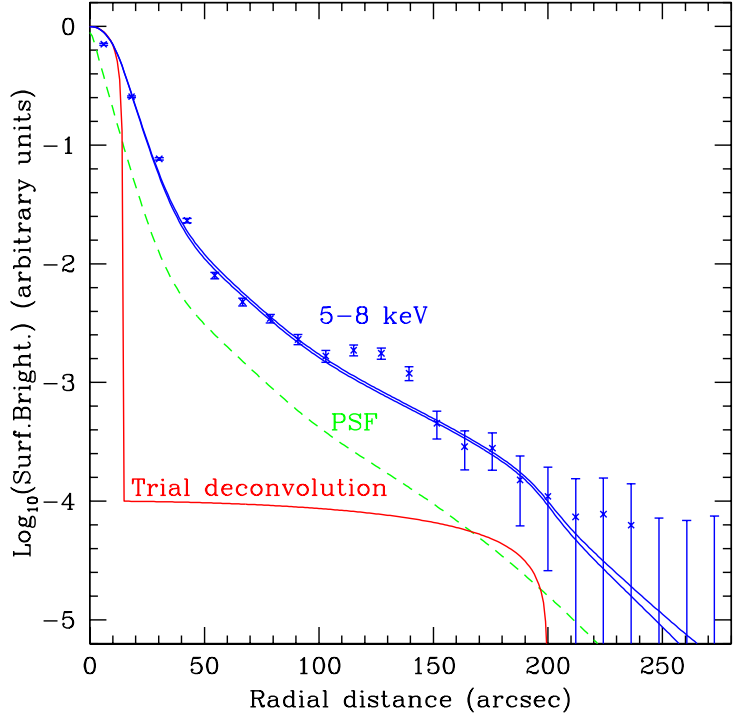}
\end{center}
\vspace{-36pt}
\caption{\ \
Residual broadening of the 5--8~keV radial profile. The data are considerably
broader than the PSF (dashed line). An intrinsic halo, modelled as a uniform
sphere of radius $200''$ and surface brightness $10^{-4}$ times the inner
component (line labelled ``Trial deconvolution''), when convolved with the PSF,
nicely fits the data. For our best-fit values, we have verified that the effect
of dust scattering in this spectral band is negligible (the 2 solid lines along
the data are respectively computed with and without dust-scattering halo). }
\end{figure}

\section*{THE NORTH SPUR}
\vspace{3pt}

A spur has been detected in the northern sector of the halo (see Figure 1, as
well as Warwick et al., 2001; Safi-Harb et al., 2001). The presence of this
intriguing feature cannot be explained by our dust model. A pure power-law fit
to the spur EPIC/PN spectrum gives acceptable $\chi^2$ values. However a pure
power-law fit to the spur gives $\NH=1.2\E{22}\U{cm^{-2}}$, namely a factor 2
less than for the other spectra: this discrepancy may indicate that a pure
power-law fit is not appropriate.

If a thermal component is included in the fit model, it is significatively
detected in the spur, while it is not detected in other halo regions. This is
shown by the (1-$\sg$, 2-$\sg$, 3-$\sg$) confidence contours plotted (Figure 5)
in the parameters plane for the thermal component: the close contours refer to
the area labelled {\it Nspur} (see Figure 1 for the areas definition), while
fits to the areas labelled {\it halo1}, {\it halo2}, and {\it halo3} give only
upper limits to the thermal fraction (open contours). In fact, adding a thermal
component in the fit to the {\it Nspur} region, the estimated $\NH$ is now
between 1.7 and $4.0\E{22}\U{cm^{-2}}$, values consistent with the other halo
regions.

\begin{figure}
\begin{center}
\includegraphics[width=0.85\textwidth]{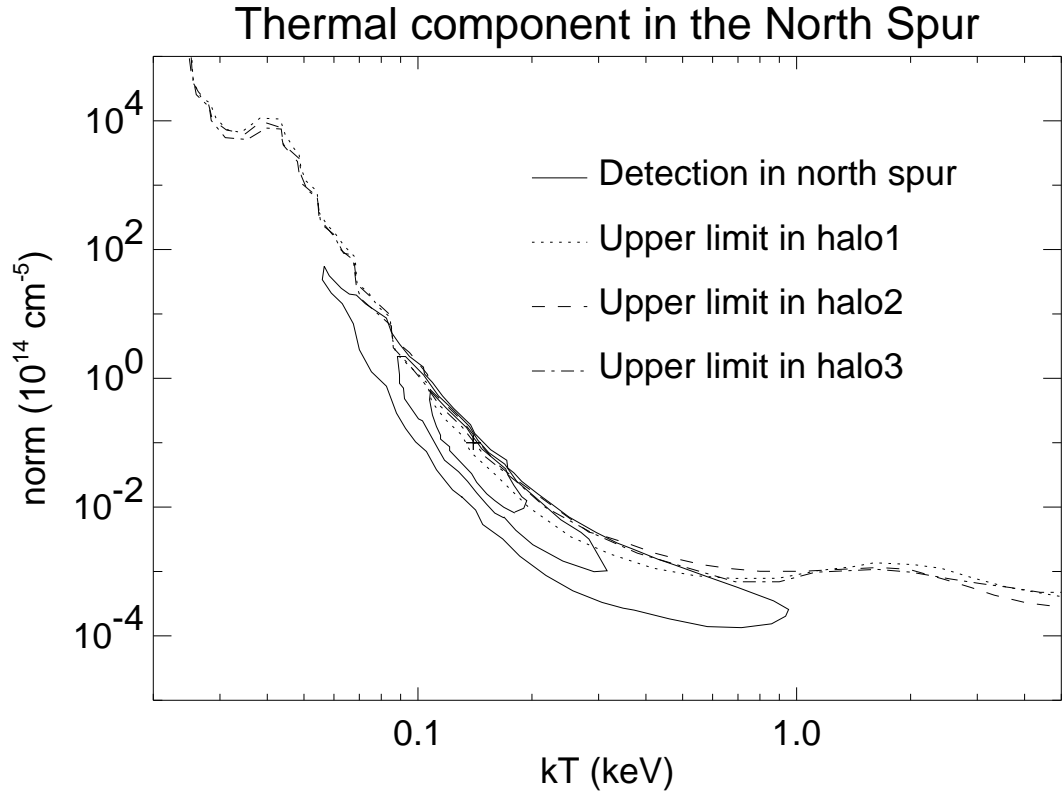}
\end{center}
\vspace{-36pt}
\caption{\ \
Confidence contours in the parameters plane for the thermal component. The
close contours refer to the {\it Nspur} region (see Figure 1); while the other
(open) contours refer to different portions of the halo.
}
\end{figure}

\vspace{6pt}
\section*{ACKNOWLEDGEMENTS}
\vspace{3pt}

This work has been partly supported by the Italian Ministry for University and
Research (MIUR) under Grant Cofin 2001--02--10.

\end{document}